\documentclass[
reprint,
amsmath,amssymb,
aps,
]{revtex4-1}
\usepackage{graphicx}
\usepackage{dcolumn}
\usepackage{bm}
\usepackage[mathlines]{lineno}
\usepackage[colorlinks=true]{hyperref}
\usepackage{mathrsfs}
\usepackage{xcolor}
\usepackage{float}
\usepackage{ulem}
\usepackage[utf8]{inputenc}

\begin{document}
\title{Interior Volume of Banados-Teitelboim-Zanelli Black Hole}
\author{Ming Zhang}
\email{mingzhang@mail.bnu.edu.cn}
\affiliation{Department of Physics, Beijing Normal University, Beijing, 100875, China}
\begin{abstract}
We calculate the interior volume for the three-dimensional Banados-Teitelboim-Zanelli black hole in this paper. To that end, we obtain the volume functional and numerically find a path to extremize it, before choosing a spacelike hypersurface to get an asymptotic analytical expression for the interior volume based on the characteristics of the path.
\end{abstract}
\maketitle

\section{Introduction}
There are two problems talking about the interior volume of a black hole: one is that time and space are interchanged inside the horizon, such that a black hole has non-static interior; the other is  how to choose spacetime slices, as the volume in general relativity depends on how spacetime is sliced. Slicing-invariant volume of a stationary black hole was defined in Ref. \cite{Parikh:2005qs}. In spacetime with a cosmological constant, the thermodynamic volume of a black hole was interpreted as a ``naive'' geometric volume conjugated to the cosmological constant in Refs. \cite{Kastor:2009wy, Cvetic:2010jb, Kubiznak:2012wp,Gunasekaran:2012dq}. A volume in terms of Kodama vector for a dynamical black hole was proposed in Ref. \cite{Hayward:1997jp}. Considering a region inside the black hole bounded by the event horizon and two distinct ingoing null cones, a definition of the volume rate for a stationary non-degenerated black hole was raised \cite{Ballik:2010rx}. The concept of ``vector volume'' was introduced by examining the rate of growth of an invariant volume for some spacetime regions along a divergence-free vector field \cite{Ballik:2013uia}.

Recently, enlightened by the daily understanding of the volume bounded by a two-dimensional sphere immersed in the flat Minkoswki spacetime, the conception of CR volume was raised \cite{Christodoulou:2014yia}. The horizon of a spherically symmetric black hole is foliated by spacelike spheres $\mathcal{S}_{v}$ labelled by the asymptotic time $v$ while the sphere $\mathcal{S}_{v}$ is defined as the one crossed by a light signal sent by a remote stationary observer at position $r$ with proper time $t=v-r$. The maximal volume of the spherically symmetric surface $\Sigma$ bounded by $\mathcal{S}_{v}$ is the so-called CR volume. It was shown that the interior volumes grow with time for the Schwarzschild black hole \cite{Christodoulou:2014yia, Bhaumik:2016sav}, Reissner-Nordstr\"{o}m (RN) black hole \cite{Christodoulou:2014yia,Ali:2018sqk}, Kerr black hole \cite{Bengtsson:2015zda,Wang:2019ake}. These results as well as the analysis of entropy of massless scalar particles inside the black holes \cite{Zhang:2015gda, Zhang:2017aqf} may provide us with implications on discussions of the information paradox, since a larger and larger volume can accommodate huger and huger information \cite{Rovelli:2017mzl}.

It is meaningful to generalize the calculation of CR volume from the asymptotic flat black hole to the anti-de Sitter (AdS) black hole, though there has been an existing investigation mainly focusing on the relation between the CR volume and the horizon area for the RN-AdS black hole \cite{Ong:2015tua}. Due to the anti-de Sitter-conformal field theory (AdS-CFT) correspondence, the information loss paradox has obtained new explanation \cite{Lowe:1999pk}; in this regard, the volume of the AdS black hole may provide some meaningful results. Besides, the investigation of the AdS black hole may give implications to the research of the Complexity-Volume (CV) duality \cite{susskind2016computational}. In consideration of these motivations, we will calculate the interior volume of the Banados-Teitelboim-Zanelli (BTZ) black hole in this paper, which is three-dimensional and asymptotically AdS. In Sec. \ref{vol}, we will give a brief review for the BTZ black hole, and then calculate its interior volume numerically and analytically. In Sec. \ref{dis}, we will discuss the implications of our result. Sec. \ref{con} will be devoted to our conclusion.

\section{The interior volume of the BTZ black hole}\label{vol}
\subsection{A brief review for the BTZ black hole}

The action is 
\begin{equation}
I=\frac{1}{2\pi}\int d^3 x \sqrt{-g}(R+2l^{-2}),
\end{equation}
where $l$ is the AdS radius relating to the cosmological constant $\Lambda$ as $-\Lambda=l^{-2}$. The factor $(16\pi G)^{-1}$ in front of the action is taken to be $(2\pi)^{-1}$ as the gravitational constant $G$ is set to be $1/8$. The BTZ black hole solution is (we set $l=\hbar=c=1$) \cite{Banados:1992wn}
\begin{equation}\label{metric}
ds^{2}=-f(r)dt^{2}+\frac{dr^{2}}{f(r)}+r^{2}d\phi^{2},
\end{equation} 
where
\begin{equation}
f(r)=-m+r^{2},
\end{equation}
and $m$ is the mass of the black hole. The horizon $r_{+}$ can be obtained from the constraint that the time-time component $f(r)=0$ and reads
\begin{equation}\label{radius}
r_{+} =\sqrt{m}.
\end{equation}
The Bekenstein-Hawking entropy $S$, Hawking temperature $T$ are
\begin{equation}\label{entropy}
S=\frac{2\pi r_{+}}{4\hbar G}=4\pi r_{+},
\end{equation}
\begin{equation}
T=\frac{r_{+}}{2\pi}.
\end{equation}
The vacuum state
\begin{equation}
ds_{vac}^{2}=-\frac{r^{2}}{l^{2}}dt^{2}+\frac{l^{2}}{r^{2}}dr^{2}+r^{2}d\phi^{2}
\end{equation}
 is equivalent to letting $m\rightarrow$ 0. As $m$ grows negative, the spacetime becomes unphysical, while the singularity and the horizon disappear. However, when $m=-1$, it appears as an AdS space with the line element
 \begin{equation}
 ds^{2}=-\left(1+\frac{r^{2}}{l^{2}}\right)dt^{2}+\left(1+\frac{l^{2}}{r^{2}}\right)dr^{2}+r^{2}d\phi^{2}.
 \end{equation}

\subsection{The formulation of the interior volume}
The geometry of  BTZ black holes can be described in Eddington-Finkelstein coordinate as
\begin{equation}\label{ef}
ds^{2}=-f(r)dv^{2}+2dvdr+r^{2}d\phi^{2},
\end{equation}
where
\begin{equation}
  v=t+\int\frac{dr}{f(r)}=t+\frac{1}{2\sqrt{m}}\log \Big{|}\frac{m-\sqrt{m} r}{m+\sqrt{m} r}\Big{|}.
\end{equation}
The metric has a Killing vector $\xi_{v}$ corresponding to the translation in $v$. It can describe the collapse matter as the coordinates are analytically continued for all $r>0$. The coordinate $v$ tends to positive infinity when $r$ decreases to 0. The coordinate $r$ is the affine parameter of ingoing null geodesics which {}has constant advanced time $v$.
 
Now we are confronted with the concept of the sphere $S_v$, which is defined as the one crossed by a light-like signal originated from the distance $r$ at time $t$. What we will study is the maximal volume of the two-dimensional surface $\Sigma$ bounded by $S_v$. 

The two-dimensional symmetric surface $\Sigma$ is in fact a direct product of a one-sphere $\mathcal{S}$ and a curve $\gamma$ in the $v-r$ plane, as 
\begin{equation}
\Sigma\equiv \gamma \times \mathcal{S},~~\gamma\mapsto [v(\lambda),r({\lambda})],
\end{equation}
where $\lambda$ is a parameter. The initial and final endpoints $\lambda_i$ and $\lambda_f$ can be set as
\begin{equation}
\gamma_i =[v(\lambda),~r(\lambda)]|_{\lambda=0}=(v_i, r_+),\label{v2m}
\end{equation}
\begin{equation}
\gamma_f =[v(\lambda),~r(\lambda)]|_{\lambda=\lambda_f}=(v , 0)\label{vf0}
\end{equation}
respectively.

Then the induced metric on the surface $\Sigma$ can be expressed as
\begin{equation}
ds_{\Sigma}^{2}=(-f(r)\dot{v}^2 +2 \dot{v}\dot{r})d\lambda^{2}+r^{2}d\phi^{2},
\end{equation}
where $\dot{v}\equiv dv/d\lambda$, $\dot{r}\equiv dr/d\lambda$. As a result, the proper volume of $\Sigma$ can be expressed as
\begin{equation}\label{sipidp}
V_{\Sigma}=\int_0^{2\pi}d\phi\int_0^{\lambda_f}\sqrt{r^2 [-f(r)\dot{v}^2 +2 \dot{v}\dot{r}]}d\lambda.
\end{equation}

\subsection{A numerical result for the interior volume}
From Eq. (\ref{sipidp}), we can see that the maximal volume  of $\Sigma$ corresponds to a proper curve $\gamma$ that extremizes the integral. According to the Langrangian mechanics, we then make effects to find the equations of motion for the Langrangian 
\begin{equation}
L(r,\dot{r},v,\dot{v})=\sqrt{r^2 [-f(r)\dot{v}^2 +2 \dot{v}\dot{r}]}.
\end{equation}
According to 
\begin{equation}
\frac{\partial L}{\partial v}-\frac{d}{d\lambda}\frac{\partial L}{\partial \dot{v}}=0,
\end{equation}
we can obtain
\begin{equation}\label{trc}
-r^2 [f(r)\dot{v}-\dot{r}]=c,
\end{equation}
where $c$ is a constant. With the extremization, we can set $L(r,\dot{r},v,\dot{v})=1$. As a result, we have 
\begin{equation}\label{vr}
r^2 [-f(r)\dot{v}^2 +2 \dot{v}\dot{r}]=1.
\end{equation}
Using Eqs. (\ref{trc}) and (\ref{vr}), we have
\begin{equation}
\dot{r}=-\frac{\sqrt{c^2+r^2f(r)}}{r^2},\label{frr2s}
\end{equation}
\begin{equation}
\dot{v}=\frac{1}{c+r^2 \dot{r}}\label{cr2dot}.
\end{equation}

The following calculation needs us to determine the maximal-volume surface $\Sigma_m$ by solving Eqs. (\ref{frr2s}) and (\ref{cr2dot}). After integrating Eq. (\ref{frr2s}) and considering the Eqs. (\ref{v2m}) and (\ref{vf0}), we can obtain
\begin{equation}\label{sqrtc2}
\lambda_f=-\int_{r_+}^{r} \frac{r^2 dr}{\sqrt{c^2 +r^2 f(r)}}.
\end{equation}

We know from Eqs. (\ref{v2m}) and (\ref{vf0}) that $r$ is dependent of $\lambda$, which means $c^2\neq -r^2 f(r)$ in Eq. (\ref{frr2s}) and $c^2>-r^2 f(r)\geqslant m^2/4\equiv c_0^2$. 

\begin{figure*}[!htbp] 
   \centering
   \includegraphics[width=7.15in]{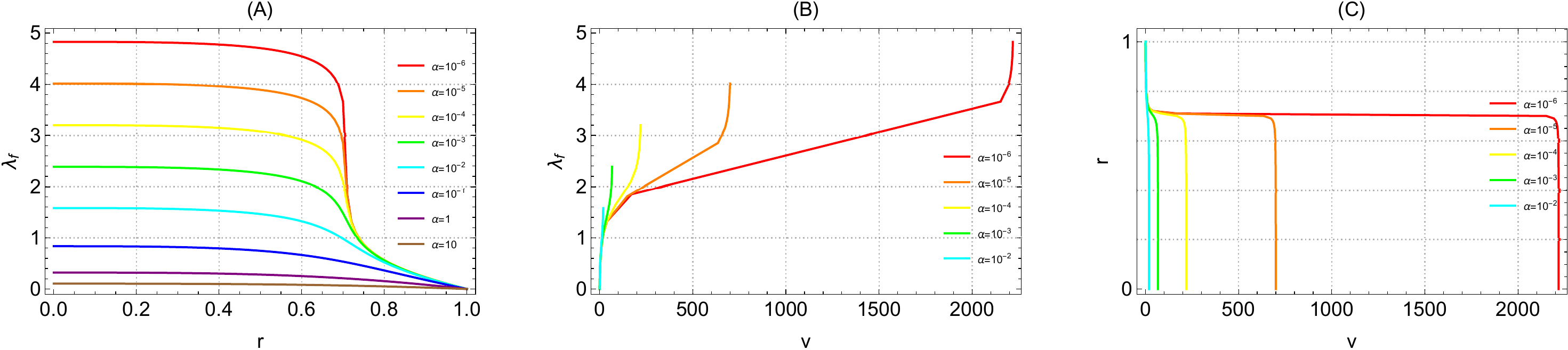}
   \caption{(A), (B): The parameter $\lambda_f$ as a function of the coordinate $r$ and $v$ for $c^2=c_0^2+\alpha,~m=1$. (C): The parameter $\lambda$ as a function of the coordinate $v$ for $c^2=c_0^2+\alpha,~m=1$.}
   \label{lambdadiagram}
\end{figure*}

We can obtain 
\begin{equation}
v-v_i=\int_{r_+}^r \frac{dr}{\dot{r}(c+r^2 \dot{r})}
\end{equation}
from Eq. (\ref{cr2dot}) as $\dot{v}=\dot{r}dv/dr$. Setting $v_i =0$ at the collapse time and using Eq. (\ref{frr2s}), we get
\begin{equation}\label{sqlettt}
v=\int_{r_{+}}^r \frac{r^2 dr}{\sqrt{c^2+r^2 f(r)}\left(-c+\sqrt{c^2+r^2 f(r)}\right)}.
\end{equation}

Combining Eqs. (\ref{sipidp}) and (\ref{vr}), we can know that 
\begin{equation}\label{sqlet}
V_\Sigma =2\pi \lambda_f.
\end{equation}
In order to find a proper path $\gamma_p$ letting $V_\Sigma$ approach its maximal value $V_{\Sigma_m}$, we can calculate $r$ and $v$ numerically using Eqs. (\ref{sqrtc2}) and (\ref{sqlettt}), and then obtain the $\lambda_f -r$, $\lambda_f -v$, and $r-v$ diagrams in Fig. \ref{lambdadiagram}.

From Fig. \ref{lambdadiagram}(A), we can see that: (1) The total $\lambda_f$ increases when $c^2$ becomes closer and closer to $c_0^2$. (2) The major value of $\lambda_f$ builds up near $r=r_c=\sqrt{m/2}$. However, $\lambda_f$ increases slowly at the beginning $(r\sim r_+)$ and the end $(r\sim 0)$. From Figs. \ref{lambdadiagram}(B) and \ref{lambdadiagram}(C), we can see that the major value of $\lambda$ builds up at $v\gg m$.

\subsection{An asymptotic analytical expression for the interior volume}
We now want to get an asymptotic analytic expression for the volume based on the numerical result in Fig. \ref{lambdadiagram}. As shown in the figure, the main contribution to the volume originates from $r\sim r_c$, we therefore can choose a hypersurface $r=r_0$ ($r_0$ is a constant and its value can be obtained once we extremize the volume) in Eqs. (\ref{frr2s}) and (\ref{cr2dot}). As a result, we can obtain
\begin{equation}
c^2=-r^2 f(r)\label{cws},
\end{equation}
\begin{equation}
\dot{v}=\frac{1}{c}\label{frla1c}.
\end{equation}
Then we can get ($v_i=0$ as we have set in above)
\begin{equation}
\lambda_f=c v
\end{equation}
from Eq. (\ref{frla1c}). Combining it with Eqs. (\ref{sqlet}) and (\ref{cws}), we can finally obtain the proper volume of $\Sigma$ as
\begin{equation}
V_{\Sigma}=2\pi c \lambda_f =2\pi\sqrt{-r^2 f(r)}v\leqslant \frac{\pi}{2}mv=V_{\Sigma_m}.
\end{equation}
The equality relation is established on the condition that $r=r_c=\sqrt{m/2}$.

In fact, once the spacelike hypersurface $r=r_0$ is chosen, we can define a restricted function $\Phi=r-r_0$, then the normal covector is 
\begin{equation}
n_{a}=\frac{\Phi_{,a}}{\sqrt{\Phi_{,\mu}\Phi_{,\nu}g^{\mu\nu}}},
\end{equation}
the corresponding normal vector is 
\begin{equation}
n^{a}=g^{ab}n_{b}.
\end{equation}
Then we can obtain the induced metric on the hypersurface as
\begin{equation}
h^{ab}=g^{ab}+n^{a}n^{b},
\end{equation}
with its determinant
\begin{equation}
h=-gN^{-2},
\end{equation}
where 
\begin{equation}
N=\frac{1}{\sqrt{\Phi_{,\mu}\Phi_{,\nu}g^{\mu\nu}}}.
\end{equation}
The proper volume of the hypersurface described by the induced metric is
\begin{equation}\begin{aligned}
V_\Sigma &=\int \sqrt{h}dvd\phi=\int\sqrt{g^{rr}g}dvd\phi=\int\sqrt{r^{2}\left|f\right|}dvd\phi\\&=2\pi v\sqrt{r^{2}\left|M-r^{2}\right|}\leqslant \frac{\pi}{2}mv.
\end{aligned}\end{equation}
In short, neglecting the contribution of the transient region near $r=r_+$ and $r=0$, we can analytically obtain the asymptotically maximal interior volume for the BTZ black hole as
\begin{equation}\label{ggmsimv}
V_{\Sigma_m}(r\sim \sqrt{\frac{m}{2}}, v\gg m)\sim 1.57 m v.
\end{equation}

\section{{}Discussions}\label{dis}
We have studied the interior volume for a spherically symmetric BTZ black hole. We first obtained the volume funtional and then tried to extremize it using numerical and analytical calculation. It should be noted that the calculation of the asymptotic volume is based on some analysis on the numerical result. Specifically, we obtained the volume of the BTZ black hole in terms of the parameter $\lambda$, together with relations $r-\lambda$, $v-\lambda$, and $r-v$. By the characteristics of the numerical data shown in the diagrams, we can know that the major contribution to the volume of the BTZ black hole originates from a specific constant $r$ hypersurface, which is related to a large advanced time $v$.

It may be interesting to calculate the volume of a rotating black hole along this way. For a rotating Kerr black hole (see Ref. \cite{a2004relativist}), one may choose an undetermined hypersurface $r=r(v,\theta)$ at first, and then insert
\begin{equation}
dr=r_v dv+r_\theta d\theta,
\end{equation}
where 
\begin{equation}
r_v\equiv \frac{\partial r}{\partial v},~r_\theta\equiv \frac{\partial r}{\partial \theta},
\end{equation}
into the Kerr metric. As a result, we can obtain the induced metric on the hypersurface $r=r(v,\theta)$. Then the volume functional together with the equations of motion can be got. One can solve the equations of motion numerically and then choose a proper hypersurface which contributes most to the volume. With the chosen hypersurface, one can then analytically calculate the asymptotic expression for the interior volume of the Kerr black hole. However, more numerical technologies should be used to solve the equations of motion and we do not plan to conduct it here.

In terms of physical significance, firstly, we extend the concept of interior volume to lower spacetime dimensions, comparing to those in higher dimensions \cite{Bhaumik:2016sav,Ong:2015tua}. 

Secondly, our result may give new insights on the problem of information paradox in the background of three-dimensional AdS spacetime. The question of information loss paradox was raised in consideration of the black hole radiation due to quantum effects \cite{Hawking:1974sw} in 1975, following with many years' discussions on this question \cite{Susskind:1993if,Almheiri:2012rt,Almheiri:2013hfa,Yan:2016vns}. The AdS-CFT correspondence \cite{maldacena1998jm,maldacena1999large,witten1998ads,gubser1998ss}, which states that there is a remarkable duality between the string theory in AdS spacetime and the CFT on the spatially infinite boundary of the AdS spacetime, provides an effective solution to the paradox \cite{Lowe:1999pk,Braunstein:2009my,Giddings:2012bm,Nomura:2012cx}. The CFT is unitary, so information of a black hole in AdS spacetime must be conserved \cite{Hawking:2005kf}.

By calculating the interior volume of the BTZ black hole in the AdS spacetime, we here touch on the information loss parodox and provide a seemingly feasible explanation to this problem besides the AdS-CFT scheme. According to the interpretations in Refs. \cite{Christodoulou:2014yia,Rovelli:2017mzl,Ong:2015dja,Christodoulou:2016tuu}, there will be larger and larger interior space to store the information, though evaporation leads to the decreasing of the horizon area.

To see this explicitly for the BTZ black hole in the AdS spacetime, we can specify the advanced time $v$ by considering the Stefan-Boltzmann law in 3-dimensional spacetime which reads \cite{landsberg1989stefan}
\begin{equation}
\frac{\text{d}M}{\text{d}v}=-\gamma AT^{3}.
\end{equation}
Here $A$ is the surface area of the radiating body (which is related to the entropy of the black hole Eq. (\ref{entropy}) by $A=4S$) and $\gamma$ is a positive Stefan-Boltzmann constant. It is valid when the mass of black hole is greater than Planck mass. Considering an evaporation process from initial mass $m$ to final mass $m_f$ (which is beyond the Planck mass), one can obtain
\begin{equation}
v=-\int_{m}^{m_f}\frac{dm^{\prime}}{4\gamma ST^{3}}\sim \frac{1}{m_f}-\frac{1}{m}.
\end{equation}
Substituting it into Eq. (\ref{ggmsimv}), we can obtain
\begin{equation}
V_{\Sigma_m}(r\sim \sqrt{\frac{m}{2}}, v\gg m)\sim \frac{m}{m_f}.
\end{equation}

It is evident that the interior volume of the BTZ black hole is almost inversly proportional to the residuary mass during the evaporation and there indeed will be more and more internal space to store the information, though the number of the available orthogonal states relating to the horizon area $A$ of the black hole, counting as $e^{A}$, will decrease with the Hawking radiation. Thus, the Hawking radiation can be corelated with the increasing internal states \cite{Rovelli:2017mzl}, though the available states relating to the black hole horizon decrease \cite{Page:1993wv}. In this respect, the consequence is consistent with those in Refs. \cite{Zhang:2015gda,Ali:2018sqk,Choudhury:2017yqh,Majhi:2017tab,Wang:2018dvo,Yang:2018arj,Dutta:2018jqc,Han2018,Wang:2018txl}, where the entropy of the massless scalar field inside the black hole was calculated using statistical thermodynamics method.

What is more, the volume of the BTZ black hole may contain interesting physics for the discussion on the {}CV duality of the {}AdS black hole \cite{susskind2016computational,Stanford:2014jda,Brown:2015lvg,Wang:2017zfn,Qaemmaqami:2017lzs}. From Eq. (\ref{ggmsimv}), we have (we set $l=1$ in the preceding text but recover it here)
\begin{equation}
\frac{\text{d}V_{\Sigma_m}}{\text{d}v}=\frac{\pi}{2}ml.
\end{equation}
In the infinite-time limit, we can view $V_{\Sigma_m}$ as the volume $V$ of the Einstein-Rosen bridge \cite{Wang:2017zfn}. Then, according to the CV duality, namely
\begin{equation}
C=k \frac{V}{l},
\end{equation}
where $C$ is the computational complexity of the instantaneous boundary CFT state and $k$ is the proportional constant, which means the volume of a certain maximal spacelike slice extending into the interior of the black hole is related to the computational complexity of the boundary CFT state  by a proportion relationship \cite{Brown:2015bva}, we have
\begin{equation}\label{cv}
\frac{\text{d}C}{\text{d}v}=2m
\end{equation}
if we set $k=4/\pi$. Remarkably, the relation (\ref{cv}) reflecting the growth of the complexity based on the CV duality is consistent to the result based on the complexity-action duality for all uncharged nonrotating AdS black holes in any spacetime dimensions \cite{Brown:2015lvg,Brown:2015bva,Cai:2017sjv,Jiang:2018pfk,Fan:2018wnv,Goto:2018iay}.

The three-dimensional AdS BTZ spacetime can be embedded into a four-dimensional flat space with the metric 
\begin{equation}
ds^2=-dT^2 -dU^2 +dX^2 +dY^2.
\end{equation}
The equation
\begin{equation}
-T^2 -U^2 +X^2 +Y^2=-l^2
\end{equation}
describing a hypersurface with constant negative curvature is a solution to the Einstein equation
\begin{equation}
R_{ab}-\frac{1}{2}g_{ab}(R-2\Lambda)=\pi T_{ab},
\end{equation}
with $\Lambda<0$ and $T_{ab}=0$. As long as we choose the local coordinates \cite{padmanabhan2010gravitation}
\begin{equation}
T=\sqrt{\frac{r^2}{m}-l^2}\sinh \frac{\sqrt{m}}{l}t,
\end{equation}
\begin{equation}
U=\frac{r}{\sqrt{m}}\cosh(\sqrt{m}\phi),
\end{equation}
\begin{equation}
X=\sqrt{\frac{r^2}{m}-l^2}\cosh \frac{\sqrt{m}}{l}t,
\end{equation}
\begin{equation}
Y=\frac{r}{\sqrt{m}}\sinh(\sqrt{m}\phi),
\end{equation}
we can obtain the metric (\ref{metric}). 
However, one should notice that the global topology between BTZ solution and AdS space is different, as we  have to view $\phi$ and $\phi +2\pi$ identically \cite{padmanabhan2010gravitation}. Our calculation of the interior volume of the BTZ black hole and discussions about the result thus are to some extent qualitatively different from the research of general AdS black holes. This, in turn, partly makes sense of our investigations in this paper.

\section{Conclusion}\label{con}
The interior of the black hole is dynamical and time-dependent. In this paper, we followed the definition of the black hole's interior volume proposed in Ref. \cite{Christodoulou:2014yia} and extended it to the volume of the three-dimensional BTZ black hole. The volume functional Eq. (\ref{sipidp}) was obtained before the extremization of it. The interior volume as the maximum value of the volume functional was calculated numerically and analytically. We obtained the maximal volume surface and showed it in Fig. \ref{lambdadiagram}. Then we chose a constant radius hypersurface to deduce the asymptotic expression for the interior volume after we discovered that the main contribution to the volume comes from a region near a certain value of the coordinate $r$. Finally, as shown in Eq. (\ref{ggmsimv}), we found that the interior volume of a BTZ black hole is proportional to its mass $m$ and asymptotic time $v$. {}Our result may give inspirations to the volume calculation of rotating black hole, the information loss paradox, as well as the CV duality of the AdS black hole.
 
\section*{Acknowledgements}
This work is supported by the National Natural Science Foundation of China (Grant No. 11235003). I would like to thank Xin-Yang Wang for helpful discussions.

\end{document}